\documentstyle[prl,aps,epsf]{revtex}
\begin{document}
\twocolumn[\hsize\textwidth\columnwidth\hsize\csname@twocolumnfalse%
\endcsname
\preprint{SU-ITP \# 97/34}
\title{ Proximity Effect and Josephson Coupling in the $SO(5)$ Theory of
  High-Tc Superconductivity }
\author{ E. Demler$^1$, A.J. Berlinsky$^1$, C. Kallin$^1$, G.B. Arnold$^2$,
  and M.R. Beasley$^3$ }
\address{\it {}$^1$ Department of Physics, Stanford University, Stanford
  CA~~94305}
\address{\it {}$^2$ Department of
  Physics, University of Notre Dame, Notre Dame, Indiana 46556 }
\address{\it {}$^3$ Department of Applied Physics, Stanford University,
  Stanford CA~~94305}
\maketitle

\begin{abstract} We consider proximity effect coupling in
Superconducting/Antiferromagnetic/Superconducting (S-A-S)
sandwiches using the recently developed $SO(5)$ effective theory of high
temperature superconductivity. We find that, for narrow junctions, the A
region acts like a strong superconductor, and that there is a critical
junction thickness which depends on the effective $SO(5)$ coupling constants
and on the phase difference across the junction, at which the A region
undergoes a Freedericksz-like transition to a state which is intermediate
between superconductor and antiferromagnet. For thick junctions, the
current-phase relation is 
sinusoidal, as in standard S-N-S and S-I-S junctions, but for thin
junctions it 
shows a sharp break in slope at the Freedericksz point. 
\end{abstract}

]

\newpage

Zhang has recently developed a theory\cite{SCZhang} which unifies d-wave
superconductivity (S) and antiferromagnetism (A) on the basis of an
underlying $SO(5)$ symmetry.  The S and A order parameters are combined
into a 5-dimensional superspin, and the high energy physics of these
superspins is postulated to be rotationally symmetric. At low energies
this $SO(5)$ symmetry is broken by a chemical-potential-dependent anisotropy
which favors the A state for $\mu < \mu_c$ or the S state for $\mu >
\mu_c $. This implies that, at low temperature, there is a ``soft
direction'' for perturbations of a stable d-wave superconductor toward
antiferromagnetism.  Similarly, the appropriate perturbation, applied to a
stable A material, will tend to drive it into the S state.  By analogy to
the proximity effect in conventional superconductors, it is clear that the
relevant perturbing field is provided by proximity of an A material to an
S material. Moreover, in a sandwich S-A-S configuration, this proximity
effect would be expected to provide a mechanism for Josephson coupling
of the two S regions. We also note that one approach to practical high-Tc
Josephson junctions involves the use of barriers made from the cuprates
near the S/A transition.

\begin{figure*}[h]
\centerline{\epsfxsize=8cm 
\epsfbox{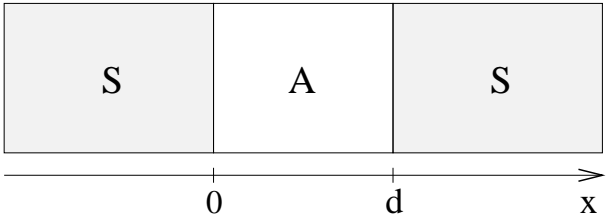}
}
\caption{Geometry of the suggested junction}
\label{geometry}
\end{figure*}

In this paper we present analytic and numerical results for the properties
of the S-A-S Josephson junction system, shown on Fig.\ref{geometry}, in
terms of $SO(5)$ continuum 
theory in which the spatial variation of the order parameter is one
dimensional.  We obtain analytical results for the critical current as
a function of thickness and 
numerical results for the current-phase relation for different
thicknesses. 

We find that, when the S layers are strongly superconducting, thin A
layers are driven completely superconducting by the field of the
adjacent S layers, and the $SO(5)$ order parameter lies completely in
the superconducting plane. Beyond a critical barrier thickness, we
find that the order parameter in the junction starts to tip back
toward the antiferromagnetic plane, in a fashion precisely analogous to the 
Freedericksz transition in liquid crystals. Twisting the
superconducting phase, which causes a current to flow through the
junction, is analogous to twisting the nematic director at the walls.
A sufficiently large twist will drive the system through the
Freedericksz transition resulting in a distinctive, non-sinusoidal
current-phase relation for an S-A-S junction.

Our results clearly demonstrate that, within $SO(5)$ theory,
the details of Josephson coupling through an A barrier are
qualitatively different from those of proximity effect junctions with
conventional barriers. Hence study of S-A-S junctions provides a
critical test of $SO(5)$ theory. By the same token our
calculations provide a new basis for the interpretation of real
high-Tc Josephson junctions, currently being fabricated and
studied\cite{suzuki,barner,hashimoto}.

\begin{figure*}[h]
\centerline{\epsfysize=4cm
\epsfbox{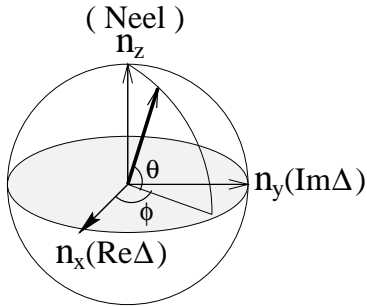}
}
\caption{$SO(5)$ order parameter  }
\label{sphere}
\end{figure*}

In the spirit of $SO(5)$ we
describe the system by a three-component order parameter ${\bf
  n} = \{ n_x, n_y, n_z\}$, where the first two components are the real
and imaginary parts of the superconducting order parameter and the third
component represents the antiferromagnetic Neel vector ( See Figure
\ref{sphere} ). For
simplicity we 
treat the Neel vector as a single component. However this component may be
viewed as the amplitude of a spatially uniform 3D vector.

According to \cite{SCZhang} the system is described by a functional
\begin{eqnarray} 
{\cal L} ({\bf n}) = \frac{\rho}{2} \left( \partial_\mu n_a \right)^2 - g
 n_z^2 \end{eqnarray}
 with the constraint $n^2 = 1$.
As in \cite{arovas} we assume that the gradient term is $SO(5)$
 symmetric. The anisotropy term $g$ is positive
in the A region (so that it would be antiferromagnetic in the absence of
proximity effects) and negative in the superconductor. 

 The superspin
constraint is most naturally implemented in polar coordinates $n_x =
\cos \theta \cos \phi$, $n_y = \cos \theta \sin \phi$, and $ n_z = \sin
\theta$. \begin{eqnarray} {\cal L} (\theta, \phi) = \frac{\rho}{2} \{
(\partial_\mu \theta )^2 + \cos^2 \theta~ ( \partial_\mu \phi )^2 \} - g~
\sin^2 \theta \label{angle-functional} \end{eqnarray} 
In all of our
 calculations we will assume rigid superconducting boundary conditions $n_z
 |_0 = 0 $, and $n_z |_d = 0 $. Strictly speaking this is only true in the
 case of ``strong'' 
superconductors and ``weak'' antiferromagnets:  $| g_{S} | \gg |g_{A}
|$. However analysis of the general case shows that relaxing this
 condition does not change the qualitative picture. 

At this point one can
easily specify the analogy between our problem and the problem of a liquid
crystal in a slab with anchoring walls, in an electric
field. If the electric field is perpendicular to the walls, it will try
to align the director of the liquid crystal along the field. At small
voltages the field is unable to overcome the effect of surface pinning,
and the equilibrium configuration remains uniform. However with
increasing voltage the system will undergo a Freedericksz transition, in
which the director begins to align along the field. More interestingly
this transition is known to depend on the applied boundary conditions,
i.e. on the relative twist of the anchoring directions on the two sides of
the slab (the twisted nematic transition) \cite{helium}.

We now show that similar effects arise in S-A-S sandwiches within
$SO(5)$ theory. The role of the voltage is played by $d \sqrt{g_{A}/\rho}$,
and the superconducting phase difference across the junction corresponds
to the twist angle imposed by the two anchoring walls. 
The S-A-S
sandwiches will undergo a phase transition in which the A region, between the
two superconductors, goes from being purely superconducting (by virtue of
the proximity effect)  into a mixed S/A state. We also show that,
sufficiently close to such a Freedericksz transition, the system possesses
non-trivial current-phase characteristics, as a consequence of the
transition.

In the A region the Euler-Lagrange equations for the functional
(\ref{angle-functional}) are 
\begin{eqnarray}
\rho \frac{d^2 \theta}{d x^2} &+& \rho \cos \theta \sin \theta \left(
  \frac{d \phi}{d x} \right)^2 +2 g_A \sin \theta \cos \theta =0
\label{euler-lagrange1} \\
\frac{d}{d x}& ( &  \cos^2 \theta \frac{d \phi}{d x}~ ) =0
\label{euler-lagrange2} 
\end{eqnarray}
The boundary conditions for these equations are given by 
\begin{eqnarray}
\theta(x=0)&=&0 \hspace{1cm} \phi(x=0)=0 \\
\theta(x=d)&=&0 \hspace{1cm} \phi(x=d)= \Delta \Phi
\end{eqnarray}
where $\Delta \Phi$ is the phase difference between two superconductors.

Equation (\ref{euler-lagrange2}) is nothing but the conservation of current.
$$I_s = n_1 \partial_x n_2 - n_2 \partial_x n_1= \cos^2 \theta \frac{d
  \phi}{d x}$$ So we can write (\ref{euler-lagrange1}) as 
\begin{eqnarray}
\rho \frac{d^2 \theta}{d x^2} + \rho \sin \theta \frac{ I_s^2} { \cos^4
  \theta} + 2 g_A  \sin \theta \cos \theta =0 
\label{euler-lagrange3} 
\end{eqnarray}
The last equation can be easily integrated once giving
\begin{eqnarray}
\xi_A^2 \left( \frac{ d \theta}{ d x} \right)^2 = -\frac{ I_s^2 \xi_A^2}{
  \cos^2 \theta } - \sin^2 \theta + \frac{ I_s^2 \xi_A^2}{
  \cos^2 \theta_0 } + \sin^2 \theta_0 
\label{eq1}
\end{eqnarray}
with the characteristic length 
\begin{eqnarray}
\xi_A = \sqrt{ \rho / 2 g_A}
\label{xi-A}
\end{eqnarray}
In writing (\ref{eq1}) we expressed the constant of integration in
terms of the maximal value $\theta_0$ that will be reached at $x = d/2
$ (where $d \theta / dx =0 $).
This immediately results in an equation for $\theta_0$
\begin{eqnarray}
\frac{d}{2 \xi_A} &=& \int_0^{\theta_0} \frac{ d \theta}{ \sqrt{ 
-\frac{ \omega_s^2 }{
  \cos^2 \theta } - \sin^2 \theta + \frac{ \omega_s^2}{
  \cos^2 \theta_0 } + \sin^2 \theta_0 
}} = \nonumber\\
&=&\frac{ \cos \theta_0 }{ \sqrt{ \omega_s^2 + \cos^2 \theta_0 }}
K(k) 
\label{d-eq}
\end{eqnarray}
where $\omega_s = I_s \xi_A$, the parameter $k$ is defined by 
\begin{eqnarray} 
k^2 = \frac{ \sin^2 \theta_0 \cos^2
  \theta_0 }{ \omega_s^2 + \cos^2 \theta_0}
\label{k-eq}
\end{eqnarray}
and K is the complete elliptic integral of the first kind.
Equation (\ref{d-eq}) should be supplemented by an equation 
for the current $\omega_s$ in terms of the phase
difference across the junction
\begin{eqnarray}
\Delta \Phi &=& 2 I_s \int_0^{d/2} \frac{d x}{ \cos^2 \theta(x) }
= 2 I_s \int_0^{\theta_0} \left( \frac{d \theta}{ d x} \right)^{-1}
\frac{ d \theta }{ \cos^2 \theta } 
\nonumber\\
&=& 2 \omega_s 
\int_0^{\theta_0} \frac{ d \theta}{ \cos^2 \theta \sqrt{    
-\frac{ \omega_s^2 }{
  \cos^2 \theta } - \sin^2 \theta + \frac{ \omega_s^2}{
  \cos^2 \theta_0 } + \sin^2 \theta_0 
}} \nonumber\\
&=& 2 \omega_s \frac{ \cos \theta_0 }{ \sqrt{ \omega_s^2 + \cos^2
    \theta_0 }} \Pi_1( -\sin^2 \theta, k )
\end{eqnarray}
here $ \Pi_1( n, k )$ is a complete elliptic integral of the third
kind.

One can easily see that Eq. (\ref{d-eq}) has a solution only
when $ d/ \xi_A \geq 
\pi / \sqrt{ 1 + \omega_s^2} $. For smaller $d$ the only solution will be
$\theta_0 = 0$, which 
means that the A region remains uniformly superconducting. Even
though antiferromagnetism would be favored in a bulk material of
this kind, proximity to a ``strong'' superconductor forces it to be
uniformly superconducting. When $d_c = \pi \xi_A / \sqrt{ 1 + \omega_s^2
  } $, a second order 
transition occurs at which $\theta_0$ starts to increase 
as $\sqrt{d-d_c}$, so that the A region exhibits both kinds of order:
superconductivity and antiferromagnetism. It is interesting to note that
a non-zero $\omega_s$ decreases the critical width of the A region.  This 
can be understood as the result of having an extra ``torque'' in the x-y
plane. This result raises the very interesting possibility of choosing a width
of the A region below the critical value at zero current $d_{c0}
= \pi \xi$ and then tuning the system through the transition by simply 
passing a current through the junction!

\begin{figure*}[h]
\centerline{\epsfysize=8cm
\epsfbox{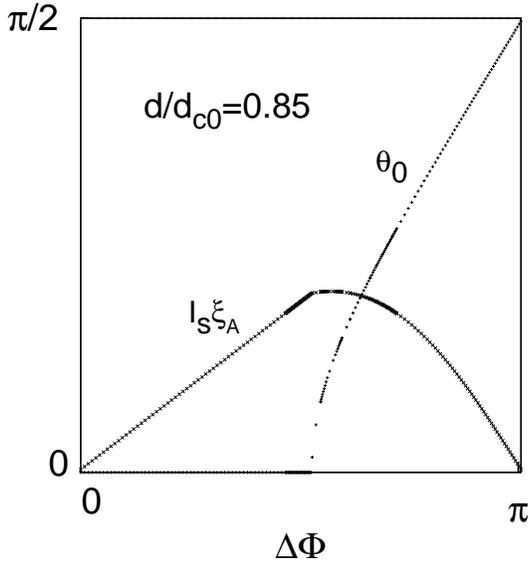}
}
\caption{$\theta_0$ vs $\Delta \phi$ and $I_s$ vs $\Delta \phi$ for
an S-A-S junction.  The onset of $\theta_0$ and the discontinuity 
in slope of $I_s$ both occur at the Freedericksz transition. }
\label{comb-fig}
\end{figure*}

In Figure \ref{comb-fig} we present such an example, for the case $d=
0.85 d_{c0}$. 
This figure shows that the system undergoes a transition when
$\Delta \Phi = 
1.7$. Below the transition $\theta_0$ is identically zero and $I_s$ is a
linear function of $\Delta \Phi$, as one would expect for a uniform
superconductor. However above the Freedericksz transition, $\theta_0$
starts to grow and $I_s$ vs $\Delta \Phi$ develops
curvature. Eventually, at $\Delta \Phi = 
\pi$, $\theta_0 = \pi/2$ and $I_s$ goes to zero. We note that
further interesting differences with the conventional proximity effect can
be expected in the dynamical state at finite voltages. In the presence
of a finite voltage across the junction, the full $SO(5)$ order
parameter will undergo periodic motion in $SO(5)$ space, permitting exploration of the low $q$-vector dynamics of $SO(5)$ theory.   

\begin{figure*}[h]
\centerline{\epsfysize=8cm
\epsfbox{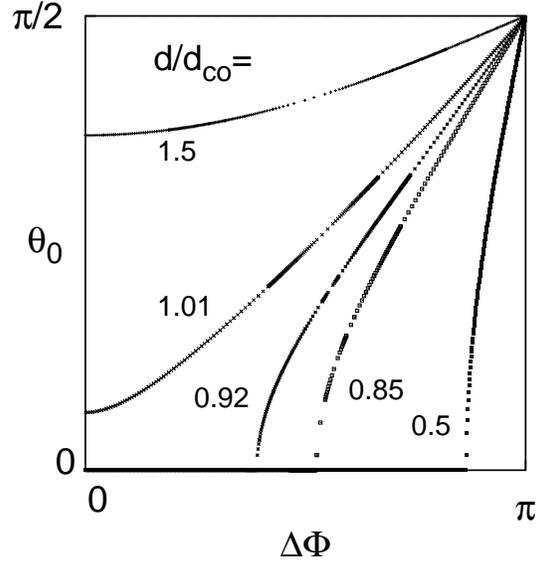}
}
\caption{ $\theta_0$ vs $\Delta \Phi$ for junctions with different
$d/d_{c0}$. Notice 
that for $\Delta \Phi = \pi$ we always have $\theta_0 = 
\pi/2$.  } 
\label{theta-fig}
\end{figure*}

Figure \ref{theta-fig} shows that this feature, $\theta_0 = 
\pi/2$ when $\Delta \Phi = \pi$, occurs for all widths of
the A region. It may be understood as follows: The energy required to
twist the superconducting order parameter by $\pi$ without changing
its magnitude is the same as the energy required to
rotate the superspin into the
antiferromagnetic plane and 
back into the superconducting plane. However rotating the superspin into
the antiferromagnetic direction allows the system to lower its energy
because of the 
$g$-term. 

\begin{figure*}[h]
\centerline{\epsfysize=8cm
\epsfbox{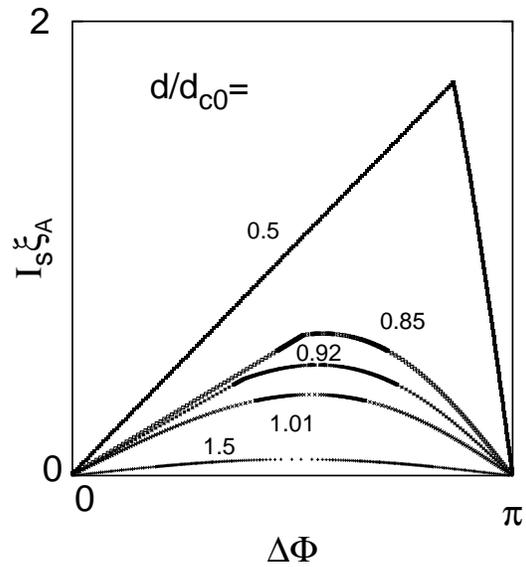}
}
\caption{ Current-phase characteristics of junctions with
different $d/d_{c0}$.  }
\label{I-fig}
\end{figure*}

This effect is an interesting $SO(5)$ analogue of
the result of Krotov {\it et. al.}\cite{Krotov} that superconductivity
between antiferromagnetic stripes is suppressed for nontopological
stripes and enhanced for topological
stripes. 

Figure \ref{I-fig} illustrates the non-trivial current-phase 
characteristics of S-A-S junctions with increasing width of the A
layer. When $d<d_{c0}$ they show a transition from linear dependence
below the transition to $\sin$-like dependence above it. Some
asymmetry persists in the curves  for $d \geq d_{co}$, and for 
$d \gg d_{c0}$ they show the usual $\sin(\Delta \Phi)$ dependences of
SIS junctions.

It is easy to calculate the critical current of our junctions. For a given $d$, Eq. (\ref{d-eq}) does not have any solution
for currents that are too large. The first solution appears at a
point that corresponds to the maximum of $k^2$ in Eq.
(\ref{k-eq}) \cite{footnote}. This $k_{max}$ is given by $$
k^2_{max}=1-2 \omega_s ( \sqrt{1+\omega_s^2} - \omega_s ) \simeq 1-2
\omega_s $$ Using the asymptotic forms of the elliptic functions, we find
for Eq. (\ref{d-eq}) $d/(2 \xi_A) = ln(4/\sqrt{2 \omega_{s}})$ which
gives the critical current
\begin{eqnarray} 
I_{s} = \frac{8}{\xi_A}~ e^{-d/{\xi_A}}. 
\end{eqnarray} 
So $\xi_A$ represents a new correlation
length for superconducting proximity effects across antiferromagnets: 
According to \cite{SCZhang} $g_{A} = 2 \chi ( 
\mu_c^2 - \mu^2 )$, where $\mu$ is the chemical potential, and $\mu_c$ is
the critical value of the chemical potential at which the first
order transition between the superconducting and antiferromagnetic
states occurs.  In deriving Eq. (\ref{xi-A}), we have
assumed an $SO(5)$ symmetric susceptibility $\chi$ for the charge,
spin and $\pi$ operators. According to Eq. (\ref{xi-A}), when $\mu$ is close
to $\mu_c$ (and hence $g_A$ is small), $\xi_A$ will be large. This
provides a new and natural explanation of  
the long range proximity effect sometimes observed in $PBCO$
\cite{suzuki,barner,hashimoto}. We note, however, that
asymmetry in the $\chi$'s will generate
a cut-off for $\xi_A$  in Eq. (\ref{xi-A}). For $\chi_c > \chi_{\pi}$ we find
$\xi_{max} = \sqrt{\rho / \eta}$ where $\eta = 2 \mu_c^2 ( \chi_c -
\chi_{\pi}) $.

The authors are greatly indebted to S.C. Zhang for numerous illuminating
discussions on the $SO(5)$ theory and for generously sharing of his ideas.
We would like to thank A. Fetter for drawing our attention to
\cite{helium}. 
A.J.B.  and C.K. acknowledge support by the Natural Sciences and
Engineering Research Council of Canada and by the Ontario Centre for
Materials Research. C.K. also acknowledges the support of a Guggenheim
Fellowship. This work is also supported by the Office of Naval
Research and the National Science Foundation through the NSF-MRSC
program.


\begin{thebibliography}{10}
\bibitem{footnote} This simple argument applies only when $w_s$ is
much smaller than one, i.e. well above the Freedericksz transition. For
smaller widths one can expect small corrections due to the prefactor of
the elliptic function in (\ref{d-eq}).
\end{thebibliography}
\end{document}